\documentstyle[spie]{article} 
\input{psfig}   
\input epsf

\title{Atomistics of Tensile Failure in Fused Silica: \\ Weakest Link Models
Revisited}

\author{J. I. Katz\supit{a} 
\skiplinehalf 
\supit{a}Department of Physics and McDonnell Center for the Space Sciences
\\ Washington University, St. Louis, Mo. 63130 USA \\ and \\ MITRE Corp.,
1820 Dolley Madison Blvd., McLean, Va. 22102 USA}

\authorinfo{Further author information: E-mail: katz@wuphys.wustl.edu}

\begin{document} 
  \maketitle 

\begin{abstract}
In weakest link models the failure of a single microscopic element of a
brittle material causes the failure of an entire macroscopic specimen, just
as a chain fails if one link fails.  Pristine samples of glass, such as
optical communications fiber, approach their ideal strength, and their
brittle tensile failure has been described by this model.  The statistics of 
weakest link models are calculable in terms of the statistics of the
individual links, which, unfortunately, are poorly known.  Use of the
skewness of the failure distribution may permit simultaneous determination
of the statistics of the individual weak links and of their number density,
which indicates their physical origin.  However, the applicability of
weakest link models to real materials remains unproven.
\end{abstract}


\section{INTRODUCTION}
\label{sect:intro}  

Leonardo da Vinci knew that longer wires fail at lower tensile loads than
shorter ones.  For a simple chain this result is so obvious it hardly bears
mention: a  chain is only as strong as its weakest link; if any link fails
the chain fails.  Adding links to a chain will weaken it if one of the
added links is weaker than any of the initial links, or will leave its
strength unchanged if all added links are stronger than the weakest initial
link.  Adding links to a given chain will inevitably weaken it if enough
links are added, if the strengths of the links are drawn randomly from a
non-singular distribution.  The weakest link found in a large sample (a long
chain) will {\it on average} be weaker than the weakest link found in a
small sample (a short chain).

These results are obvious to us, but may have been less obvious in
Leonardo's day, before the invention of the science of mechanics (and
statistics), when the mechanics of a chain was imperfectly
understood.  In fact, it is not completely trivial, for it depends on the
loading conditions, the properties of the material of which the links are
made, and the definition of failure.  For example, if the links are capable
of work-hardening and failure is defined as extension of the chain by a
suitable fixed length, then a long chain may actually be able (depending on
its detailed properties) to sustain a greater load than a short one.
\section{WEAKEST LINK MODELS}
From the picture of a chain in tension a simple idealized model, known as
the weakest link model, is derived.  In this model a fiber (wire, filament
or rod) loaded in tension is described as a series of independent mechanical
elements ordered in a single file, each of which transmits a load (force, or
momentum flux) from its immediate neighbor on one side to its immediate
neighbor on the other side.  The model is quasi-static: all elements are
assumed to be subjected to the same tensile load, which will be the case if
there are no accelerations and no forces on the links except those exerted
by their immediate neighbors.  It is also local, in the sense that force is
transmitted only across the boundaries between two elements; this assumption
may fail if the elements are coupled by long range forces, such as
electromagnetism, as in a current-carrying or charged chain or one made of
piezoelectric material, although these effects are unlikely to be
significant in practice.  Failure of the entire chain is defined as the
failure of a single element.  Given these assumptions, the strength of a
chain is the strength of its weakest link.

The statistics of weakest link models were apparently first considered by
Peirce\cite{Peirce26}, who worked for the British Cotton Industry Research
Association and published in the Journal of the Textile Institute.  This is
curious because weakest link models are unlikely to be applicable to ropes
and cables made up of many parallel strands.  These strands divide the load
amongst themselves, so that the failure of a single strand need not, and
usually does not, imply the failure of the entire rope.  We are all familiar
with ropes that are partly frayed, but which still carry a load.
Multi-stranded ropes are also non-local, because the shift of load from a
failed strand to the remaining strands occurs by static friction over an
extended length of the rope, and depends in a complex manner on the lay of
the rope.  In addition, load shifting occurs as a result of differential
initial length, elastic stiffness and creep between strands, even when there
has been no failure.

The statistics of weakest link models were reviewed by Epstein\cite{Epstein48},
whose paper remains useful today.  He made a number of significant points: 
\begin{enumerate}
\item If the links have a statistical distribution of strengths, then the
mean strength of the entire chain (the strength of the weakest link) is a
decreasing, albeit generally very slowly decreasing, function of the number
of links in the chain.
\item The distribution of chain strengths is generally a skew function, with
a longer ``tail'' in the direction of weaker chains; this is a consequence
of the fact that a single unusually weak link makes a weak chain, while an
unusually strong link has no effect on the strength of the chain.
\item For normally distributed link strengths, the dispersion in chain
strengths is a slowly decreasing function of the number of links.
\item In many physically important problems the weakest link model is
inapplicable; he cites notches on the surface of a bent beam as an example.
A single notch serves as a site of stress concentration, while a large
number of closely spaced notches (a roughened surface), if of equal depth,
reduce the stress in the matter below them to low values, thus reducing the
stress at the apex of any one notch.
\end{enumerate}

Weakest link models are expected to be applicable to the tensile failure of
brittle materials in many, but not all, circumstances.  The reason for this,
known since the work of Griffith\cite{Griffith20}, is that brittle materials
usually fail in tension because of stress concentration at
imperfections---surface scratches, voids, or heterogeneities in the bulk.
Once local failure occurs, the size of the imperfection and its degree of
stress concentration increase, and the object ruptures.  

Suppose a macroscopic brittle object may be divided into small cells, each
of which contains no more than one imperfection, and each imperfection is
contained within a single cell.  In most materials these cells must be very
small because the density of imperfections is large, but in pristine glass
fiber the imperfections may be so rare that a cell can encompass the entire
diameter of the fiber and extend in length for many diameters.  If such a
division into cells is possible, with negligible interaction between the
different cells and the imperfections within them, then we would expect a
weakest link model to be applicable, because the failure of any one cell
would lead to the failure of the entire object.  This is true even though
the cells need not be arrayed in a line like links in a chain; even cells
side-by-side are independent potential sites of failure.

An essential assumption is that imperfections do not interact elastically.
This is valid if the imperfections are small and well separated in all their
dimensions, but may not hold if they are geometrically extended, such as
grain boundaries in a ceramic or surface scratches.  In such more complex
geometries it may again be possible to use weakest link models if the cells
are large enough to contain many imperfections, for then each cell may act
as an independent link, even though the individual imperfections interact
strongly.  In such a case, the resulting statistics will be those of the
cells, not those of the individual imperfections.

It is a curious feature of models of this kind that a renormalization group
transformation is not possible (their statistics are not invariant under
such a transformation) because the statistics of weakest link models depend
on the total number of links, which is not invariant.  This follows from the
assumption that the flaws have a characteristic (small) size, and are not
``fractal''.  It also assumes that a single local failure is sufficient to
produce general failure; in other words, that general failure is not a
consequence of the interaction of many failures in the type of
process\cite{Katz86} now known as self-organized criticality\cite{Bak96}.

The dependence of the statistics of weakest link models on the number of
participating links means that it may be possible to learn something about
the microphysics of failure from the statistics of the strength of
macroscopic objects.  That is the subject of this paper.


  \section{FUSED SILICA} 

Fused silica and related silica glasses are materials of enormous
technological importance.  They are also among the few materials (the others
are carbon nanotubes and, when creep is negligible, metallic whiskers) which
may approach their ideal limiting strengths because they can be produced
apparently without volume or surface stress-concentrating flaws.
\subsection{History}
Griffith\cite{Griffith20} was apparently the first to realize the importance
of flaws in reducing the strength of brittle materials.
Griffith also found that finer glass fibers were systematically stronger than
thicker ones, which may naturally be attributed (in a weakest link model) to
the presence of more links in the thicker fibers.  He compared this result
to the similar results of Karmarsch from 1858 on metal wires, although we
would now consider brittle (glass) and ductile (metallic) failure to be so
different that a close comparison is not justified.  For historical reviews
see the work of Holloway\cite{Holloway85} and Kurkjian\cite{Kurkjian85}.

Griffith measured tensile strengths under ordinary room conditions of short
lengths of very fine (3--4$\,\mu$ diameter) glass fibers of about 500,000 psi,
which are comparable to those measured under similar conditions today.  He
also noted the effect of aging and of static fatigue in reducing these
strengths, which remain important subjects of research.  He speculated
that the strength of these fine fibers could be extrapolated to
infinitesimal diameters to determine the limiting strength of ideal glass.
This extrapolation is inconsistent with his hypothesis that he had produced
flaw-free samples and not very meaningful because it was based on data
obtained under ambient conditions for aged samples; we now know that ideal
strength requires an inert environment.  Despite this, his numerical value
($1.6 \times 10^6$ psi) for the ideal strength of glass is quite close to
modern values.

\begin{table} [h]   
\caption{Griffith's data on strength of glass fibers} 
\label{tab:Griffith}
\begin{center}       
\begin{tabular}{|c|c|c|c|} 
\hline
\rule[-1ex]{0pt}{3.5ex} &&& \\
\rule[-1ex]{0pt}{3.5ex}  Diameter & Breaking Stress & Diameter & Breaking
Stress \\
\rule[-1ex]{0pt}{3.5ex} &&& \\
\hline
\rule[-1ex]{0pt}{3.5ex} &&& \\
\rule[-1ex]{0pt}{3.5ex}  0.001 inch & lbs. per sq. inch & 0.001 inch & lbs.
per sq. inch \\
\rule[-1ex]{0pt}{3.5ex} &&& \\
\rule[-1ex]{0pt}{3.5ex}  40.00 & 24,900 & 0.95 & 117,000  \\
\rule[-1ex]{0pt}{3.5ex}  4.20 & 42,300 & 0.75 & 134,000  \\
\rule[-1ex]{0pt}{3.5ex}  2.78 & 50,800 & 0.70 & 164,000  \\
\rule[-1ex]{0pt}{3.5ex}  2.25 & 64,100 & 0.60 & 185,000  \\
\rule[-1ex]{0pt}{3.5ex}  2.00 & 79,600 & 0.56 & 154,000  \\
\rule[-1ex]{0pt}{3.5ex}  1.85 & 88,500 & 0.50 & 195,000  \\
\rule[-1ex]{0pt}{3.5ex}  1.75 & 82,600 & 0.38 & 232,000  \\
\rule[-1ex]{0pt}{3.5ex}  1.40 & 85,200 & 0.26 & 332,000  \\
\rule[-1ex]{0pt}{3.5ex}  1.32 & 99,500 & 0.165 & 498,000  \\
\rule[-1ex]{0pt}{3.5ex}  1.15 & 88,700 & 0.130 & 491,000  \\
\rule[-1ex]{0pt}{3.5ex} &&& \\
\hline 
\end{tabular}
\end{center}
\end{table}
Some of Griffith's data are also surprising, given modern understanding.  He
found (Table 1) a fairly smooth dependence of strength on diameter, while we
would expect fibers to be either without mechanical damage and very strong,
or damaged and very weak; finer fibers would be expected to be more
frequently found undamaged, but not systematically stronger than undamaged
thicker fibers.  We would also expect the strength to be essentially
independent of diameter for the finest (rarely damaged) fibers, so that no
extrapolation to zero diameter would be necessary or appropriate.  After the
lapse of 80 years it is impossible to interpret all his data, but some
of his results may be a consequence of his testing of aged samples and his
use of fiber diameter as an independent variable, in contrast to modern
experiments which generally use communications fiber of a single diameter.


At the time of Griffith's work no theory, reasonable in modern terms, of the
ideal strength of materials existed.  Approximate theories were soon
developed.  Polanyi\cite{Polanyi21} used a simple energetic argument to
predict the ideal strength of crystals from the energy required to form new
surfaces by rupture.  Although he did not refer to the (one year earlier) 
work of Griffith\cite{Griffith20}, his surface energy-based argument
resembles Griffith's argument for the growth of a crack, differing chiefly
in that Griffith considered (using Inglis's analytic solution) stress
concentration at the crack tip while Polanyi made the na\"\i ve assumption
of a cleavage with plane parallel faces.  From this Polanyi correctly
concluded that for most materials the measured strength is far below any
reasonable theoretical estimate, but did not point out stress concentration
as the explanation.
\subsection{Properties}
The development of quantum mechanics and the theory of interatomic forces
made possible more quantitative models of the theoretical strength of
materials, beginning with the work of Frenkel\cite{Frenkel26}.  Modern band
structure theory makes reliable predictions of ideal strengths possible for
crystalline materials, although these are generally unobservable because
their much lower real strengths are determined by stress concentration at
flaws for brittle materials and by dislocation motion for ductile materials.
The modern methods are applicable to carbon nanotubes, which may perhaps be
made free of defects because they are so small.  They may also undergo plastic
flow\cite{Zhang98}, which may prevent them from reaching their ideal
strength, even if free of defects, and they are known to buckle reversibly
on bending.
 
   %

Fused silica and silica glasses have apparently not been the subject of
modern theoretical strength calculations, probably because it is difficult
to specify their non-crystalline structure.  This structure can be
calculated by molecular dynamics methods, but the result is dependent on the
model for the interatomic forces and represents only a small subvolume of
the bulk material; extension to the bulk requires a questionable
extrapolation (such as periodic boundary conditions).  $\beta$-cristobalite,
a high-temperature crystalline phase of silica with cubic
symmetry\cite{Wyckoff63} and a
density close to that of fused silica, may be a useful model, but its
properties do not appear to have been calculated from first principles and
even its elastic constants have not been measured.
 
The development of fiber optic communications produced a renewal of research
on fused silica.  Essentially defect-free fiber is now available in enormous
quantity, generally encapsulated in a protective coating which prevents
mechanical damage and reduces the rate of environmental chemical attack.
The tensile strength of this material, directly measured\cite{Proctor67} at
low temperature (77$^{\,\circ}$K) where chemical attack is believed to be
negligible, may reasonably be taken to be its ideal strength, and is about
140 Kbar.  Low-temperature measurements\cite{France80} in which the stress
is inferred from the strain in two-point bending give similar results, but
are less direct because the reduction of the data depends on applying thin
beam theory to a geometry in which it is not quantitatively applicable
(because the radius of curvature is so small) and because the nonlinear
stress-strain relation of fused silica is somewhat uncertain.  The measured
low-temperature strength is about 0.2 of the Young's modulus, in accord with
rough estimates of the strengths of the chemical bonds.  Strengths measured
under ambient conditions are about one third of those measured at low
temperatures, and depend on the temperature, humidity, and duration of load,
indicating complex processes of stress corrosion by water.

\section{MICROSCOPIC PHYSICS}

Now that the theory of strength-reducing cracks is well understood, and the
strength of fused silica has been measured reasonably well, what more can be
learned?
\subsection{Mechanisms}
It took many years before the importance of surface damage, as opposed to
volume flaws, for the strength of glass was appreciated\cite{Holloway85}.
However, the strength of an unflawed (pristine) sample of glass, its ideal
strength, is likely determined in its bulk, by the strength of its network
of covalent bonds.  This does not answer the question of which component of
the glass, or which physical process, actually determines that strength; it
is possible that its strength is actually determined by the bonds at its
surface.

A number of candidate strength-limiting components or processes need be
considered.  On the largest scale, a small heterogeneity may provide a
slight degree of stress concentration and therefore determine the strength.
The extraordinarily small measured\cite{Kurkjian83} dispersion ($\ll 1$\%) of
the strength of pristine glass and the at least semi-quantitative approach of
the measured low-temperature strength to its estimated ideal value appear to
make this implausible, unless the degree of stress concentration is very
slight, as might be produced by heterogeneities of small amplitude (a small
amplitude fluctuation in elastic constants).  In optical communications
fiber the germania-doped core provides such a fluctuation, but it is ordered
rather than random and homogeneous (to good accuracy) along the length of the
fiber.

Some heterogeneity is implicit in the fact that glasses are amorphous, and
on the smallest spatial scale significant heterogeneity arises from the
fact that Si---O bonds are not all equivalent; each has a slightly different
length, and the O---Si---O and Si---O---Si bond angles differ from their
ideal values.  Thus, in a piece of glass subjected to tension in a specified
direction different bonds will fail at different stresses.

In addition to SiO$_2$, real glass, even fused silica, contains other
components.  There are a variety of impurities, reduced to very low levels
in communications-grade optical fiber, but present nonetheless.  These
introduce bonds other than Si---O bonds, and these other bonds will
generally be weaker (as is shown by the fact that mixed silicate glasses are
less strong than fused silica\cite{France80}).  Some of these impurities
(such as Na$_2$O and CaO in soda-lime glass) are present at high abundances,
while others (such as transition metal oxides in communications-grade fiber)
are very rare, but all are sites of weakness.  There may be other kinds of
strength-limiting defects, such as non-stoichiometrically bonded atoms,
interstitials, {\it etc.}.  In the core of optical communications fiber
several percent of the Si has been replaced by Ge.  Because Ge atoms are
slightly larger than Si atoms the bond network is further distorted there,
and at the boundary between the pure SiO$_2$ and the GeO$_2$-substituted
regions there is systematic strain, analogous to imperfect lattice parameter
matching in epitaxial growth of crystals.  These effects are distinct
from heterogeneity of bulk elastic constants referred to above.

Finding which of these mechanisms actually determines the strength of glass,
particularly fused silica, is of both scientific interest and practical
importance, for some mechanisms may be affected by changes in composition or
processing, potentially leading to stronger fibers.  Analogous
considerations may apply to the more complicated problem of room-temperature
strength, where the resistance to stress-assisted chemical attack determines
the strength, rather than ideal mechanical failure alone.
\subsection{Can Failure Statistics Help?}
For each of the possible modes of failure a macroscopic (or even mesoscopic)
piece of glass contains a large number of well-separated elastically
independent regions.  If the failure of a single region leads to general
failure, a plausible but unproven assumption, then a weakest link model will
apply.  The specific mechanisms of failure differ greatly in the number $N$
of independent potential failure sites present per unit volume, or in a
given sample.  For example, in a 1 m length of 125$\,\mu$ diameter fiber
the number of elastic heterogeneities or geometrical irregularities may be
very small ($N \sim 1$--$10^4$), the number of Si---O bonds is $N \sim
10^{21}$, and the number of rare impurities may (depending on the impurity)
be in the range $0 \le N < 10^{15}$.

One means of determining the sources of failure is to perform measurements
on specimens in which the number of each type of potential failure site is
varied.  This is difficult in practice, partly because it may not be
possible to control their number, and partly because each series of
experiment would require the production of a batch of fiber of custom
composition.

It is a general feature of weakest link models that as $N$ increases the
strength of the specimen (as in the classical chain or wire) decreases.  If
the statistics of the strengths of the individual links were known this
fact, by itself, would be used to determine $N$ from the dependence of
strength on size\cite{Epstein48}, but these statistics are rarely known, and
not in the present case.  It is also often difficult to perform experiments
with a sufficiently large range of specimen sizes.  Similarly, the
dispersion of strength among specimens\cite{Epstein48} depends on $N$ and
could be used to determine it, but again only if the statistics of the
individual links are known.  These two effects are closely related, and do
not give independent information.

Quantitative analysis depends on the assumed statistics of the links, and
many possible distributions have been considered.  One possible
model\cite{Katz98}, closely related to classical Weibull statistics,
describes each link by a fracture readiness parameter $\alpha$, which may be
considered a stress concentration factor, or the reciprocal of a link's
ideal strength.  The values of $\alpha$ are distributed according to a
distribution $f(\alpha)$, which is defined for $\alpha > 0$ and normalized
\begin{equation}
\int_0^\infty f(\alpha^\prime)\,d\alpha^\prime = 1.
\end{equation}
The sample fails if the largest fracture readiness parameter found among the
$N$ links, $\alpha_{max}$, exceeds a value $\alpha_0$.  The probability that
it does not fail is, to good approximation if $N \gg 1$,
\begin{equation}
P(\alpha_0) \approx \exp{\left(-\int_{\alpha_0}^\infty N f(\alpha^\prime)\,
d\alpha^\prime\right)}.
\end{equation}
The probability that failure occurs for a value of $\alpha_0$ between
$\alpha$ and $\alpha + d\alpha$ is $P(\alpha)d\alpha$, where
\begin{equation}
P(\alpha) = N f(\alpha) \exp{\left(-\int_\alpha^\infty N f(\alpha^\prime)\,
d\alpha^\prime\right)}.
\end{equation}

The most useful way to parametrize the results is as the ratio of the width
$w$ of $P(\alpha)$ to the value $\alpha_{max}$ at which $P(\alpha)$ is a
maximum; in terms of the Weibull modulus $m$
\begin{equation}
{w \over \alpha_{max}} \approx {1 \over [m(m-1)]^{1/2}},
\end{equation}
and the approximation is almost exact if $w$ is defined as the dispersion of
$P(\alpha)$
\begin{equation}
w \equiv \left(\left\vert{d^2 \ln P(\alpha) \over d\alpha^2}\right\vert_{
\alpha = \alpha_{max}}\right)^{-1/2}
\end{equation}
and $m \gg 1$.

The stretched exponential function is defined
\begin{equation}
f(\alpha) = {C(\nu) \over \alpha_0}\exp{\left[-\left({\alpha \over \alpha_0}
\right)^\nu\right]}.
\end{equation}
This is a general form widely used when the actual functional form is
unknown, and includes the simple exponential and Gaussian as special cases.
The normalizing constant $C(\nu) \equiv \nu/\Gamma(1/\nu)$ and $N^\prime
\equiv N C(\nu)/\nu$.  The distribution $P(\alpha)$ is shown in Figure 1 for
$\nu = 2$ (a Gaussian).
\begin{figure}
\begin{center}
\begin{tabular}{c}
\epsfxsize=10cm
\epsfysize=8cm
\epsfbox{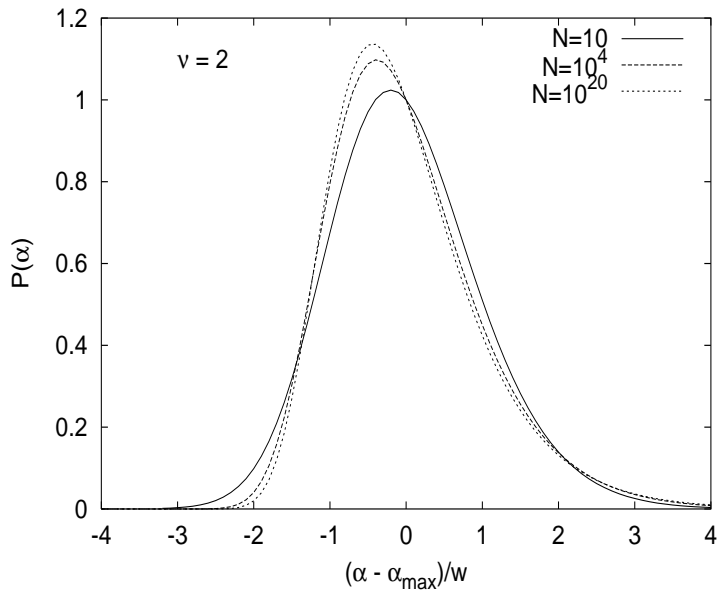}
\end{tabular}
\end{center}
\caption{Distribution $P({\alpha})$, normalized to $P(\alpha_{max})$, as a
function of $(\alpha - \alpha_{max})/w$ for a Gaussian $f(\alpha)$.
The substantial skewness is evident, as is the dependence on $N$.}
\end{figure}

By successive approximations,
\begin{equation}
{w \over \alpha_{max}} \approx {\left\{1 - {1 \over 2 \ln N^\prime} \left[
{\nu - 1 \over \nu} - \left({\nu - 1 \over \nu}\right)^2 \ln \ln N^\prime
\right]\right\} \over \nu \ln N^\prime \left(1 - {\nu - 1 \over \nu} {\ln \ln
N^\prime \over \ln N^\prime}\right)^{1/\nu}} \approx {1 \over \nu \ln
N^\prime}.
\end{equation}

It is evident that if $f(\alpha)$ is, or can be fitted to, a stretched
exponential or to one of its special cases useful and plausible estimates of
$N \approx \exp{[\alpha_{max}/(w\nu)]}$ can be obtained.  However, the
inferred value of $N$ depends very sensitively on $\nu$, and measurement of
$m$ alone for a sample of test specimens of a single size does not determine
$\nu$.  Measurement of two or more populations of very different-sized test
specimens of the same material (for which $N$ is proportional to the size)
may determine both $N$ and $\nu$, and may be feasible; for example, in
two-point bending experiments on optical fiber of 125$\,\mu$ diameter the
number of atoms $N_a$ at significant risk of initiating fracture (those
with stresses within about $1/m$ of the maximum, where the Weibull
modulus $m > 100$, which are found only close to the outside of the sharpest
part of the bend) is $N_a \sim 5 \times 10^{14}$, while in tensile loading
of a 50 m gauge length of fiber $N_a \sim 4 \times 10^{22}$ atoms are
uniformly loaded.  Two or more measurements of $m$ in which very different
numbers of atoms are stressed permit simultaneous determination of both the
ratio $N/N_a$ and $\nu$, although no extant data serve the purpose, in part
because bending and tensile measurements are affected differently by
variations in fiber diameter.
\subsection{Skewness}
In order to obtain useful results, even with data of unprecedented accuracy,
it will probably be necessary to measure more than the dispersion of the
strength.  It has long been known, both theoretically within weakest link
models\cite{Epstein48} and empirically, that distributions of failure
strengths are strongly skewed, with outliers preferentially found in the
direction of anomalously weak specimens.  The reason for this is, of course,
that an unusually weak link produces an unusually weak specimen, while an
unusually strong link has no effect at all on the specimen strength.
Quantitative measurements of the skewness of the strength distribution may
constrain the statistics, and, more importantly, the number of individual
links contributing.

The skewness of $P(\alpha)$ is defined:
\begin{equation}
s \equiv {\int_0^\infty (\alpha -\alpha_{max})^3 P(\alpha)\,d\alpha \over
w^3 \int_0^\infty P(\alpha)\,d\alpha}.
\end{equation}
The skewness is not small; see Fig. 1.  It is also not readily estimated
analytically because the Taylor expansion of $P(\alpha)$ does not converge
sufficiently rapidly, but it may be calculated numerically.  Values of the
skewness, as a function of $\nu$ and $N$, are shown in Fig. 2.
\begin{figure}
\begin{center}
\begin{tabular}{c}
\epsfxsize=10cm
\epsfysize=10cm
\epsfbox{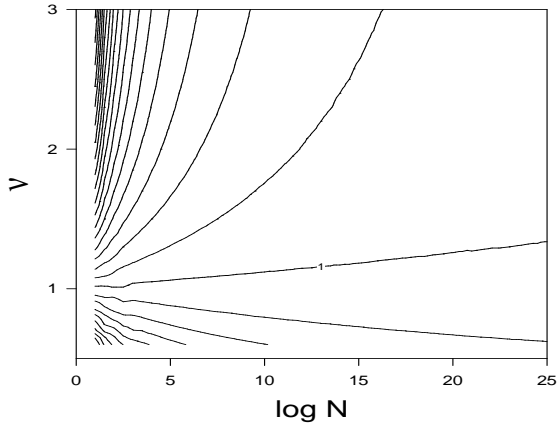}
\vspace{-2.5cm}
\end{tabular}
\end{center}
\caption{Contour graph of calculated skewnesses of $P(\alpha)$ as a function
of $\nu$ and $N$.  Contour intervals are 0.05, increasing downward.
Integrations (to calculate both skewness and dispersion) were
self-consistently truncated at $\vert \alpha - \alpha_{max} \vert = 5w$.}
\end{figure}

The values plotted were computed using cutoffs on the integrals of $\pm 5 w$
from $\alpha_{max}$, with $w$ defined self-consistently using the same
cutoff.  The reason for this is that $P(\alpha)$ has a long tail extending
toward increasing $\alpha$ which contributes significantly to the skewness,
but which is unlikely to be observed in a real experiment with a reasonable
number of specimens because there will probably be no specimens that far
out in the the tail of the distribution of strengths.  The skewness computed
without this cutoff is significantly larger, typically by $O(10\%)$.

When comparing experimental data to Fig. 2 a similar cutoff must be applied
to the data.  This will, in addition, exclude samples which are anomalously
weak because of mechanical damage or other gross flaws, which otherwise must
be excluded {\it ad hoc}.  If the number of measurements is not large it may
be necessary to choose a narrower cutoff, and to recompute Fig. 2
accordingly.
\subsection{Beyond Skewness}
Skewness is not the only statistical parameter beyond dispersion which may
be useful.  In principle, there are an infinite number of moments of the
strength distribution, but the higher moments (and the skewness, too) are
difficult to calculate reliably and accurately, even from data of high
quality, because they are very sensitive to outliers.  A data set of
reasonable size will not adequately sample rare outliers, and, in addition,
there will generally be an additional population of outliers resulting from
damaged specimens or errors of measurement.  It is difficult, and perhaps
impossible, to decide in an unbiased manner if an outlier should be
discarded as a probably consequence of a bad specimen or error, or retained
in calculating the skewness and higher moments, and even a single outlier
may have a substantial effect on their calculated values.  Therefore, it is
better to truncate the data distribution deliberately in defining the
moments (as was done for the skewness in Fig. 2).

Rather than use moments, it may be more informative (and robust) to compare
the complete distributions of strength to models.  This avoids the problem
of outliers, bur requires large data sets.  While the truncated skewness can
be determined to useful accuracy with comparatively few data (1000
measurements will typically determine it to $\pm < 0.1$ if outliers are
excluded), comparing distributions divided among many bins may require more
data, simply because with few data per bin statistical fluctuations are more
important.  However, optical communications fiber is cheap, and automated
testing machines may permit the measurement of thousands of specimens.

\section{DISCUSSION}
In order to use successfully the methods discussed here it will be necessary
to obtain strength measurements of unprecedented accuracy and number.
For example, it is believed\cite{Kurkjian83} that
the dispersion in existing measurements of the strength of fused silica
optical fiber is largely the result of dispersion in fiber diameter, rather
than dispersion in the actual strengths of the fiber.  This problem is
likely to occur whenever small dispersions in strength are measured, and may
have contributed to Leonardo's observation that long wires are weaker than
short ones: a long wire is more likely to contain a region of narrower
diameter, just as it is more likely to contain an intrinsic flaw, and it is
not straightforward to distinguish the two effects.

Equation (7) indicates that the dispersion in intrinsic strengths is
unlikely to be much less than 1\% unless the individual links have a
fractional dispersion in strength $\ll 1$.  Well controlled fused silica
fibers have a dispersion in diameter of 0.4\% and in cross-section of 0.8\%,
and an apparent dispersion in strength comparable to that in
cross-section\cite{Kurkjian83}.  In order to measure the real dispersion in
strength much tighter control of fiber diameters will be required.
These data already hint that the intrinsic fractional dispersion in strength
of individual links may be small, which gives significant information about
the nature of the weakest links---perhaps surprisingly, it points towards
the SiO$_2$ matrix, in which all chemical bonds are nearly the same, and away
from rare impurities, which might be expected to involve a more
heterogeneous range of bond strengths.  It should be remembered, however,
that these measurements were performed at room temperature, and reflect
failure by stress corrosion rather than the ideal materials strength.

Weakest link models have been widely discussed for over half a century, but
their applicability to real materials remains unproven.  In part, this is
because analysis of failure statistics rarely goes beyond calculation of the
dispersion and Weibull modulus, and plotting on Weibull coordinates.  There
are few detailed quantitative studies of failure statistics, which would
include measurement of the dependence of the dispersion and mean strength on
specimen size, as well as quantitative study of the distribution of failure
stresses.  For specimens affected by heterogeneities and damage the data are
unlikely to justify more careful analysis, but optical communications fiber
offers both the opportunity to investigate the ultimate limit of ideal
materials strength and the possibility of homogeneous data of sufficient
quality to make such an investigation possible.

In fact, there are qualitative reasons for questioning the applicability of
weakest link models to the ideal strength of glass.  As discussed, extant
data\cite{Kurkjian83} indicate a surprisingly small dispersion in measured
strength.  The surprise is increased by the fact that no sample can be
completely free of chemical impurities and radiation damage, which would be
expected to introduce anomalously weak links.  The three-dimensional network
of Si---O bonds may not be described by weakest link models, but may instead
show a well-defined collective failure limit which leads to homogeneous
macroscopic behavior from a microscopically heterogeneous
material\cite{Katz86,Bak96}.

Understanding the statistics of the ideal strength of glass at low
temperatures is of interest to the physicist.  Understanding its strength at
ambient conditions is of much greater practical interest, and nearly all
data have been obtained under these conditions.  It is also unclear if
weakest link models are applicable to stress corrosion; the simple picture
of the failure of a single small link leading to a crack propagating across
the entire specimen is not obviously applicable.  The kinetics of stress
corrosion of glass are not understood quantitatively, and the collection of
more accurate statistics of its strength distribution may be a new and
useful means of attacking that problem.

\acknowledgments     
 
I thank A. E. Carlsson for discussions and DARPA for support.

  \bibliography{glassconf}   
  \bibliographystyle{spiebib}   
 
  \end{document}